\documentclass[english,10.5pt,aps,prd,a4paper,preprintnumbers,floatfix,nofootinbib,showpacs,superscriptaddress, notitlepage]{revtex4-1} 
 \pdfoutput=1
\usepackage{color}
\usepackage{graphicx}
\usepackage{amsmath}
\usepackage{amssymb}
\usepackage{multirow}
\usepackage[colorlinks=true,citecolor=darkred,urlcolor=darkred, pdfborder={0 0 0}]{hyperref}
\usepackage[normalem]{ulem}
\definecolor{darkred}{rgb}{0,0.6,0}

\newcommand{\ba}{\begin{array}}
\newcommand{\ea}{\end{array}}
\def\be{\begin{equation}}
\def\ee{\end{equation}}
\def\bea{\begin{eqnarray}}
\def\eea{\end{eqnarray}}
\def\gsim{\ \rlap{\raise 2pt\hbox{$>$}}{\lower 2pt \hbox{$\sim$}}\ }
\def\lsim{\ \rlap{\raise 2pt\hbox{$<$}}{\lower 2pt \hbox{$\sim$}}\ }
\def\dslash{\kern-4pt \not{\hbox{\kern-2pt $\partial$}}}
\def\pslash{\not{\hbox{\kern-2pt p}}}

\newcommand{\tmm}{tetra-maximal mixing pattern}

\newcommand{\AddrUNAM}{Instituto de F\'isica, Universidad Nacional Aut\'onoma de M\'exico, A.P. 20-364, Ciudad de M\'exico 01000, M\'exico.}

\begin{document}
\DeclareGraphicsExtensions{.eps,.ps}
%
\title{\boldmath Breaking of the tetra-maximal neutrino mixing pattern}
%
%



\author{Newton Nath}\email{newton@fisica.unam.mx}\affiliation{\AddrUNAM}

\begin{abstract}
{\noindent
We make an attempt to study the present status of the tetra-maximal neutrino mixing (TMM) pattern.
It predicts all the  three leptonic mixing angles $\theta_{13} \approx 8.4^\circ,  \theta_{12} \approx 30.4^\circ, $ and $\theta_{23} = 45^\circ $ together with the three CP-violating phases $ - \delta =  \rho = \sigma = 90^\circ $. However, the latest global analysis of neutrino oscillation data  prefer relatively higher best-fit value of $ \theta_{12} $ as well as non-maximal values of both $ \theta_{23}, \delta $. In order to explain the realistic data, we study the breaking of TMM pattern.
We first examine the breaking of TMM due to renormalization group (RG) running effects and then study the impact of explicit breaking terms.
We also examine the effect of RG-induced symmetry breaking on the effective Majorana neutrino mass in  neutrinoless double beta decay experiments.
}
\end{abstract}

\maketitle

\section{Introduction}\label{sec:Intro}
The discovery of neutrino oscillations \cite{Tanabashi:2018oca} have been confirmed by various phenomenal experiments$ - $ such as solar, reactor, and, most recently, long baseline experiments. The explanation of neutrino oscillations implies the need for non-zero neutrino masses and the flavor mixing pattern of leptons.
In the Standard Model (SM) of particle physics neutrinos are supposed to be massless,  the observation of  neutrino oscillations thus hint physics beyond the SM.
%
%
%
%
The leptonic flavor mixing pattern is described by $ 3\times3 $ unitary mixing matrix $ V $ (or Pontecorvo-Maki-Nakagawa-Sakata (PMNS) matrix). 
In the standard  PDG parametrization~\cite{Tanabashi:2018oca}, the PMNS matrix $ V $ can be decomposed as
\begin{equation}\label{eq:pmns}
V = \left(\begin{matrix}c_{12} c_{13} & s_{12} c_{13} & s_{13}
e^{-i\delta} \cr -s_{12} c_{23} - c_{12} s_{23} s_{13}e^{i\delta} &
c_{12} c_{23} - s_{12} s_{23} s_{13}e^{i\delta} & s_{23} c_{13} \cr
s_{12} s_{23} - c_{12} c_{23} s_{13}e^{i\delta} & -c_{12} s_{23} -
s_{12} c_{23} s_{13}e^{i\delta} & c_{23} c_{13}
\end{matrix}\right) \left(\begin{matrix}e^{i\rho} & 0 & 0 \cr 0 &
e^{i\sigma} & 0 \cr 0 & 0 & 1\end{matrix}\right) \; ,
\end{equation}
where $s_{ij} (c_{ij}) \equiv \sin \theta_{ij} (\cos\theta_{ij})$ for $i< j = 1, 2, 3$, $ \delta $ denotes the Dirac type CP-violating phase and $ \rho, \sigma $ are the Majorana phases.
According to the latest neutrino oscillation results~\cite{Esteban:2018azc}, the best-fit values of neutrino oscillations  parameters are $\Delta m^2_{21} = 7.39 \times 10^{-5} {\rm eV}^2, ~\Delta m^2_{31} = 2.525 \times 10^{-3} {\rm eV}^2, ~ \theta_{12} = 33.82^\circ, ~\theta_{13} = 8.61^\circ,~ \theta_{23} = 48.3^\circ ,$ and $ \delta = 222^\circ$.

%

Understanding of leptonic flavor mixing pattern is still a mystifying issue in neutrino physics. 
Flavor symmetry has been very successful    in predicting  the structure of the leptonic
mixing matrix as discussed in~\cite{Altarelli:2010gt,Altarelli:2012ss,Smirnov:2011jv,Ishimori:2010au,King:2013eh}.
Among number of such symmetry bases studies, the lepton mixing matrix which is approximately equal to the tribimaximal (TBM)
mixing proposed in \cite{Harrison:2002er} turns out to be favored one. 
Considering the latest neutrino oscillation data~\cite{Esteban:2018azc,Capozzi:2016rtj,deSalas:2017kay}, the  $\mu-\tau$ reflection symmetry, which was originally proposed in Ref.~\cite{Harrison:2002et}  (see Ref.~\cite{Xing:2015fdg} for a latest review and the references therein) leads to $ |V_{\mu i }| = |V_{\tau i }|$, (for i = 1, 2, 3) as given by Eq.~\ref{eq:pmns} has received a great deal of attention in recent times \footnote{The importance of such symmetry for upcoming long baseline experiment DUNE has been studied in~\cite{Nath:2018xkz}}. 
An immediate consequence of such symmetry is that it predicts the maximal values of the atmospheric mixing angle $\theta_{23}^{} = 45^\circ$ and the Dirac type CP-phase $\delta = \pm 90^\circ$.
%
It also predicts the trivial Majorana CP-phases $ \rho, \sigma = 0^\circ, 90^\circ $. However, the mixing angles $ \theta_{13},  \theta_{12}  $ are free within this symmetry. 

Here we focus on the tetra-maximal leptonic mixing pattern, which was originally proposed by Xing in Ref.~\cite{Xing:2008ie}, in presence of latest oscillation data~\cite{Esteban:2018azc,Capozzi:2016rtj,deSalas:2017kay} \footnote{The original idea was to construct neutrino mixing pattern with only two small integers 1 and 2 together with
their square roots and the imaginary number $i$.}.
 The breaking effect of such symmetry has been examined in \cite{Zhang:2011aw}.
The tetra-maximal mixing matrix can be decomposed into four
maximal rotations
\begin{equation}\label{eq:VtmmRot}
V_0 = P_l \otimes O_{23}(\pi/4, \pi/2) \otimes
O_{13}(\pi/4,0) \otimes O_{12}(\pi/4,0) \otimes O_{13}(\pi/4,\pi) \;,
\end{equation}
where $P_l = {\rm diag}\{1, 1, i\}$, and $O_{ij}(\theta_{ij},\delta_{ij})$ is a rotation with the angle $\theta_{ij}$ and the phase $\delta_{ij}$ in the complex $i$-$j$
plane for $ij = 12, 23, 13$. The name `\textit{tetra-maximal mixing}' arises because the mixing matrix $V_0$, as given by Eq.~\ref{eq:VtmmRot}, can be expressed as a product of four rotation matrices, where all the mixing angles are maximal. 
Expanded form of  $V_0$ can be written as

\begin{equation} \label{eq:Vtmm}
V_0 = \frac{1}{2} \left(\begin{matrix} \displaystyle 1 +
\frac{1}{\sqrt{2}} & 1 & \displaystyle 1 - \frac{1}{\sqrt{2}} \cr
\displaystyle -\frac{1}{\sqrt{2}}\left[1 + i(1 -
\frac{1}{\sqrt{2}})\right] & \displaystyle 1 + i\frac{1}{\sqrt{2}} &
\displaystyle \frac{1}{\sqrt{2}}\left[1 - i(1 + \frac{1}{\sqrt{2}})\right] \cr
\displaystyle
-\frac{1}{\sqrt{2}}\left[1 - i(1 - \frac{1}{\sqrt{2}})\right] &
\displaystyle 1 - i\frac{1}{\sqrt{2}} & \displaystyle
\frac{1}{\sqrt{2}}\left[1 + i(1 + \frac{1}{\sqrt{2}})\right]
\end{matrix}\right) \; .
\end{equation}
One can extract three mixing angles as
\begin{eqnarray}
\tan \theta_{12} = 2 - \sqrt{2} \; , ~~~ \tan \theta_{23} = 1 \; ,
~~~ \sin \theta_{13} = \frac{1}{4}(2 - \sqrt{2}) \; ,
\end{eqnarray}
where $\theta_{12} \approx 30.4^\circ$,
$\theta_{13} \approx 8.4^\circ$, $\theta_{23} = 45^\circ$ and $\delta = - 90^\circ$. It also predicts Majorana CP-phases $ \rho, \sigma =  90^\circ $. 
We notice that the predicted $ \theta_{13},  \theta_{23}$ and $ \delta $ are in excellent agreement with the latest oscillation results~\cite{Esteban:2018azc,Capozzi:2016rtj,deSalas:2017kay}. However, mixing angle $ \theta_{12} $ lies out side the current 3$ \sigma $ range and the predicted value is much smaller than the latest best-fit value. 
%
%
Although the latest T2K~\cite{Abe:2017uxa} results are in good agreement with the symmetry, the current NO$\nu$A~\cite{NOVA2018} results seem to favor non-maximal $\theta_{23}^{}, \delta$.
Moreover, the latest best-fit values of the global analysis of neutrino oscillations data also   favors the same ~\cite{Esteban:2018azc,Capozzi:2016rtj,deSalas:2017kay}.

In this work, we study possible deviations from the tetra-maximal mixing pattern. We aim to explain realistic value of the solar mixing angle $ \theta_{12} $ together with non-maximal values of  the atmospheric mixing angle $\theta_{23}^{}$ and the Dirac type CP violating phase $ \delta$. In what follows we first study the breaking of tetra-maximal mixing pattern due to  the renormalization group equations (RGE).
%
As the flavor symmetries are generally imposed at a superhigh energy scale  to address neutrino masses and their flavor mixing at low energies, RGE-running
effect may lead to possible corrections and naturally break the exact symmetry.
%
%
%
Therefore, we impose  the tetra-maximal mixing  symmetry at the superhigh energy scale $ \Lambda_{\rm TMM} $ and 
analyze the deviations due to RGE breaking all the way from $ \Lambda_{\rm TMM} $ down to the electroweak (EW) scale $ \Lambda_{\rm EW} $.  Keeping the current global best-fit results in mind, which prefers $ \theta_{12} $ considerably large compared to TMM and $ \theta_{23} > 45^\circ $ and $ \delta < 270^\circ $, we study the correlation among these parameters due to RGE corrections at low energies.
Furthermore, whether neutrinos  are Majorana or Dirac
fermions, is yet unanswered in particle physics \footnote{E. Majorana first hypothesized that a fermion can be its own antiparticle in 1937~\cite{Majorana:1937vz}.}. As the symmetry predicts some trivial values for the Majorana CP-phases, which may also have some  significant impact on  neutrinoless double beta ($ 0\nu\beta \beta $) decay experiments.
Considering $ \rho, \sigma = 90^\circ $ at the energy scale $ \Lambda_{\rm TMM} $ as our  initial choice, we  study their correlation at low energies due to quantum corrections. 
We further examine the impact of RGE-induced symmetry breaking on the effective Majorana neutrino mass in $ 0\nu\beta \beta $-decay experiments.

Afterwards we perform the breaking of tetra-maximal mixing by introducing explicit breaking term in the 
leptonic mixing matrix.
The explicit breaking parameters have been introduced in such a way that one can explain large solar mixing angle and non-maximal values of atmospheric mixing angle including Dirac type CP violating phase. 
In next Sec.~(\ref{sec:RgeBr}), we examine the impact of RGE-induced symmetry breaking and perform various correlation study to explain the present neutrino oscillation data. Impact of explicit breaking of the tetra-maximal mixing has been addressed in Sec.~(\ref{sec:ExpBrm}).
Finally, we summarize our conclusion in Sec.~(\ref{sec:summary}). 
\section{Spontaneous Breaking of TMM}\label{sec:RgeBr}
We have noticed that the mixing angles $ \theta_{13}, \theta_{23} $ predicted by the \tmm~ 
are in excellent agreement with the latest neutrino oscillation data, whereas $ \theta_{12} $ lies outside $3 \sigma $ \cite{Esteban:2018azc,Capozzi:2016rtj,Esteban:2018azc}. One possible way to explain the most realistic leptonic mixing parameters under the \tmm~ is to break the symmetry. 
In general, if one assumes that the concerned mixing pattern is realized under a certain flavor symmetry at superhigh energy scale then the RGE-running will significantly impact the mixing parameters which are measured at low-energies.
Thus, the RGE-running effect provides us the possible explanation between the predicted superhigh energy scales mixing parameters to that from the experimentally measured parameters at low energies. 
In this section, we describe the 
breaking of \tmm~ due to RGE-running in context of minimal supersymmetric standard model (MSSM) \footnote{Note that the MSSM has been adopted as the theoretical framework at high energies which can serve as a possible ultraviolet extension of the Standard Model.}. 
Here, we introduce the \tmm~ at superhigh energy scale $ \Lambda_{\rm TMM} $ ($ \equiv 10^{14} $ GeV), much higher compared to electroweak (EW) scale  $ \Lambda_{\rm EW} $ ($ \sim 10^{2} $ GeV).
%

%
In Refs.~\cite{Chankowski:1993tx,*Babu:1993qv,*Antusch:2001ck,*Schmidt:2007nq,*Chakrabortty:2008zh, *Blennow:2011mp}, authors have derived the leptonic mixing parameters due to RGE-running within the numerous theoretical frameworks.
%
Considering the leading order approximation, we write down the renormalization group equations for leptonic  mixing angles in the MSSM~\cite{Antusch:2003kp} as
\begin{eqnarray}\label{eq:RGEAngles}
\dfrac{d \theta_{12}}{dt} & \approx & -\frac{y_\tau^2 s^2_{12} c^2_{12}
s_{23}^2 }{8\pi^2 \Delta m^2_{21}} \left[m^2_1 + m^2_2 + 2 m_1
m_2 \cos{2(\rho - \sigma)}\right] \;,
\nonumber \\
\dfrac{d \theta_{13}}{dt} & \approx &  \frac{y_\tau^2 s^2_{12}c^2_{12}
s^2_{23} c^2_{23} m_3}{2\pi^2\Delta m^2_{31}\left(1+\zeta\right)} \left[ m_1 \cos{(2\rho+\delta)} -
\left(1+\zeta\right) m_2 \cos{(2\sigma+\delta)} - \zeta m_3 \cos\delta
\right] \;,
\nonumber \\
\dfrac{d \theta_{23}}{dt} & \approx & -\frac{y_\tau^2 s^2_{23} c^2_{23}
}{8\pi^2\Delta m^2_{31}} \left[c_{12}^2 \left(m^2_2 + m^2_3 +
2m_2 m_3 \cos{2\sigma}\right) + \frac{s_{12}^2 \left(m^2_1 + m^2_3 + 2
m_1 m_3 \cos{2\rho}\right)}{1+\zeta} \right] \;,
\end{eqnarray}
with $t = \ln (\mu/\mu_0)$, $\zeta \equiv \Delta m^2_{21}/\Delta m^2_{31}$ 
 and $y_\tau$ denotes the Yukawa coupling of the
charged-lepton $ \tau $. 

Similarly, the different CP-violating phases can be  given by~\cite{Antusch:2003kp}
\begin{eqnarray}\label{eq:RGEPhases}
\dfrac{d \delta}{dt}  &=& \frac{y_\tau^2 s^2_{12} c^2_{12} s^2_{23} c^2_{23}
m_3 \theta^{-1}_{13}}{2\pi^2 \Delta m^2_{31}
\left(1+\zeta\right)} \left[\left(1 + \zeta\right) m_2
\sin{(2\sigma+\delta)} - m_1 \sin{(2\rho+\delta)} + \zeta m_3 \sin_\delta
\right] \;,
\nonumber \\
\dfrac{d \rho}{dt} &=& \frac{y_\tau^2}{8\pi^2} \left\{m_3
(c^2_{23}-s^2_{23}) \frac{m_1 s_{12}^2 \sin{2\rho} + \left( 1+\zeta
\right) m_2 c_{12}^2 \sin{2\sigma}}{\Delta m^2_{31} \left(
1+\zeta \right)} + \frac{m_1 m_2 c_{12}^2 s_{23}^2
\sin{2(\rho-\sigma)}}{\Delta m^2_{21}} \right\} \;,
\nonumber \\
\dfrac{d \sigma}{dt} &=& \frac{y_\tau^2}{8\pi^2} \left\{m_3
(c^2_{23}-s^2_{23}) \frac{m_1 s_{12}^2 \sin{2\rho} + \left( 1+\zeta
\right) m_2 c_{12}^2 \sin{2\sigma}}{\Delta m^2_{31} \left(
1+\zeta \right)} + \frac{m_1 m_2 s_{12}^2 s_{23}^2
\sin{2(\rho-\sigma)}}{\Delta m^2_{21}} \right\} \; ,
\end{eqnarray}
where both in Eqs.~\ref{eq:RGEAngles}, \ref{eq:RGEPhases} ${\cal O}(\theta_{13})$ term have been safely neglected.

\begin{figure}[h!]
\centering
\includegraphics[height=12cm,width=16cm]{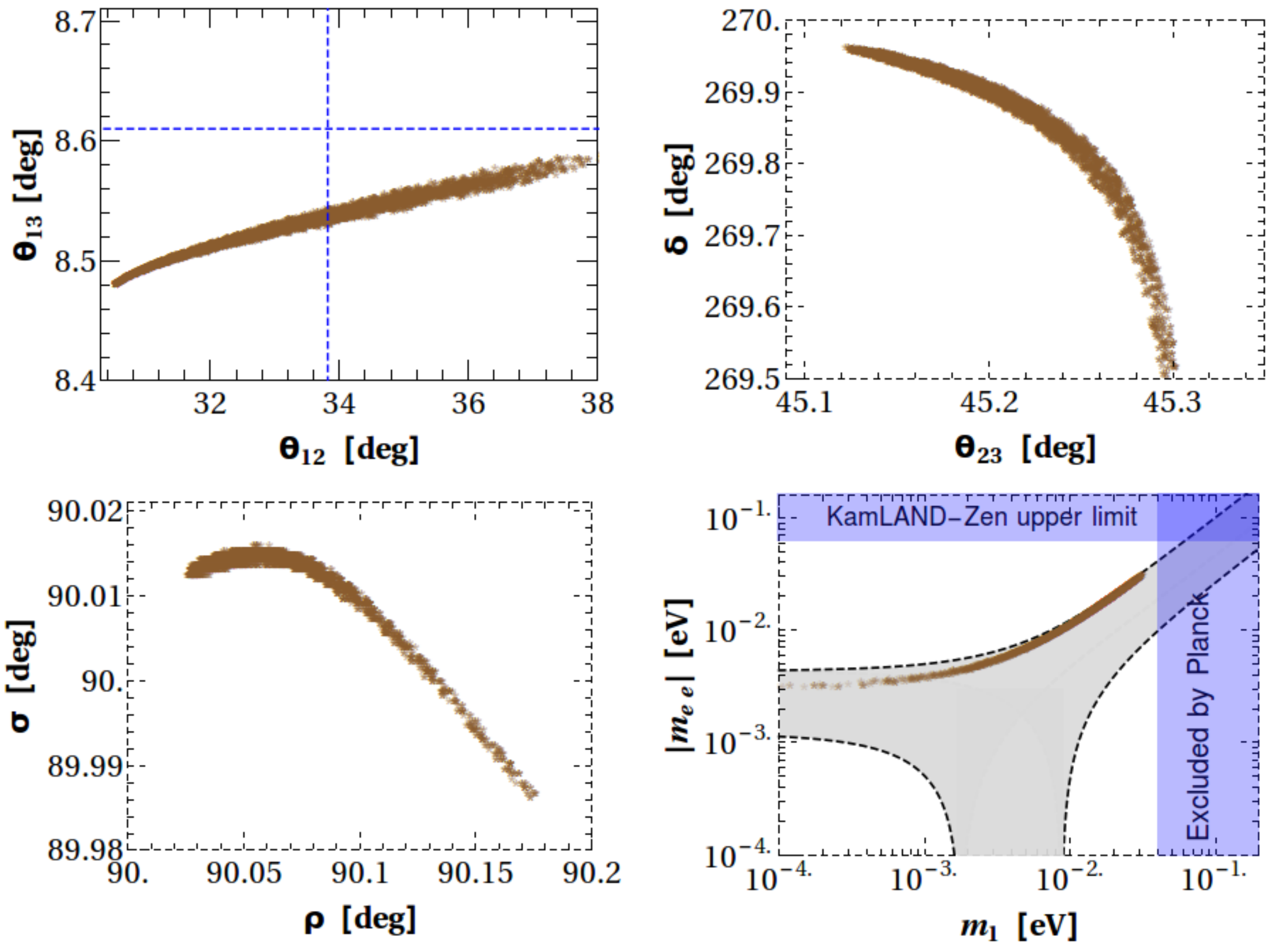}
\caption{\footnotesize Correlation plots between different neutrino oscillation parameters at $ \Lambda_{\rm EW}$. The best-fit values of $\theta_{13}$ and $\theta_{12}$ are shown by the blue dotted lines in the first panel. The fourth panel shows the prediction for the effective Majorana neutrino mass $ |m_{ee}|  $. The most stringent upper bound on $ |m_{ee}|  $ from KamLAND-Zen collaboration are shown by the light blue-horizontal band. The latest result on lightest neutrino mass is shown by the light blue-vertical band from \textit{Planck} Collaboration which gives $ \sum m_{\nu} < 0.12$ eV at the 95\% C.L. 
}
\label{fig:RGEplot}
\end{figure}

In Fig.~\ref{fig:RGEplot}, we show our numerical results  due to RGE running for the leptonic mixing parameters. 
Before moving on to discuss our results, we first illustrate the numerical procedure that has been carried out in this section. 
In the numerical study, we set the high and low energy boundary scales at $ \Lambda_{\rm TMM} = 10^{14}$ GeV and $  \Lambda_{\rm EW} = 91$ GeV together with  $ \tan \beta = 30 $, respectively. As the \tmm~ predicts  the maximal value of $ \theta_{23} $, CP-violating phases ($- \delta = \rho = \sigma = 90^\circ $) along with $ \theta_{12} \sim 30.4^\circ, \theta_{13} \sim 8.4^\circ$, we fix these values at superhigh energy scale.
Furthermore, the mass-squared differences (namely, $\Delta m^2_{31} $ and  $\Delta m^2_{31}$) are scanned over wide ranges with the help of the nested sampling package \texttt{MultiNest} program \cite{Feroz:2007kg,*Feroz:2008xx,*Feroz:2013hea} at $ \Lambda_{\rm TMM } $.
%
We define the Gaussian-$\chi^2$ function in the  numerical scan as,
\begin{equation}
\chi^{2} = \sum_i \dfrac{\left[  \xi_i -  \overline{\xi}_i \right] ^{2}  }{\sigma^2_i} \;,
\end{equation}
where $\xi_i = \lbrace \theta_{12}, \theta_{13}, \theta_{23}, \Delta m^2_{21}, \Delta m^2_{31}  \rbrace $ represents the neutrino oscillation parameters at $ \Lambda_{\rm EW} $. Also, $\overline{\xi}_i$ stands for the best-fit values from the recent global-fit results~\cite{Esteban:2018azc}, and $\sigma_i$ represents the   symmetrized 1$\sigma$ errors.
%

The best-fit values and the $1\sigma$ errors that we have considered in our numerical simulations~\cite{Esteban:2018azc} are $ \sin^2 \theta_{12} = 0.310^{+0.013}_{-0.012},~\sin^2 \theta_{13} = 0.02241^{+0.00065}_{-0.00065},~ \sin^2 \theta_{23} = 0.58^{+0.017}_{-0.021},~ \delta/^\circ = 215^{+40}_{-29},~ \Delta m^2_{21} = 7.39^{+0.21}_{-0.20} \times 10^{-5} {\rm eV}^2, ~\Delta m^2_{31} = 2.525^{+0.033}_{-0.032} \times 10^{-3} {\rm eV}^2 $. Moreover, as the latest neutrino oscillation data favors normal neutrino mass ordering (i.e., $\Delta m^2_{31} > 0$) with more than 3$ \sigma $ C.L. \cite{Esteban:2018azc,Capozzi:2016rtj,deSalas:2017kay} over 
inverted neutrino mass ordering (i.e., $ \Delta m^2_{31} < 0$), we concentrate this study considering the former mass ordering and the smallest neutrino mass $ m_1 $ is allowed to vary in the range [0, 0.2] eV. 


The results describe for mixing parameters in Fig.~\ref{fig:RGEplot} are in excellent agreement with the analytical expressions as given by Eqs.~\ref{eq:RGEAngles}, \ref{eq:RGEPhases}.
By inspecting the top row,  one can notice that the mixing angle $\theta_{12}$ receives the sizable amount of corrections due to RGE induced symmetry breaking than the remaining mixing angles $\theta_{13}$ and $\theta_{23}$. This is because of the presence of $ \zeta \approx 30 $ terms in the expression of  $\theta_{13}$ and $\theta_{23}$ (see Eq.~\ref{eq:RGEAngles}). Moreover, we find that all the mixing angles receive positive RGE corrections.
From the first panel of Fig.~\ref{fig:RGEplot}, one can observe that the mixing angle  $ \theta_{12} $ is able to achieve the latest best-fit value $\theta_{12} = 33.82^\circ$  as shown by the vertical dashed blue line. Also, shows good agreement with the latest $ 3\sigma $ data.
One can understand this behavior from Eq.~\ref{eq:RGEAngles}. Within the symmetry, imposing $ \rho = \sigma = 90^\circ  $, we find that the leading order term of $ \theta_{12} $ is $\propto (m_2 + m_1)/(m_2 - m_1) > 1$  which enhances the evolution of $ \theta_{12} $.
On the other hand, $\theta_{13}, \theta_{23}$ can not reach their current best-fit values i.e., $\theta_{13} = 8.61^\circ$ (see dashed blue horizontal line), $\theta_{23} = 49.6^\circ$ as can be seen from the left and right panels of the first row, respectively. 
A deviation of less than $\mathcal{O}(1^\circ) $ have  been identified for both the parameters.
 This is because of the suppression factor of $ \zeta $ term in their respective expressions as given by Eq.~\ref{eq:RGEAngles}.
However, both the parameters can fit the  $ 3\sigma $ global-fit data in a very efficient way ~\cite{Esteban:2018azc}.

%

Besides this, for the CP-violating phases, imposing  $  - \delta = \rho = \sigma = 90^\circ   $  in Eqs.~\ref{eq:RGEPhases} as an initial condition,  we notice from the first  line of Eqs.~\ref{eq:RGEPhases} that the leading order term of the  Dirac type CP phase $ \delta $ is $\ll 1$. Thus, we find less than $\mathcal{O}(1^\circ) $ deviation for $ \delta $ from the second panel of  Fig.~\ref{fig:RGEplot}. The mild deviation that can be seen from the figure is due to the higher order terms. 
From the second and third lines of Eqs.~\ref{eq:RGEPhases}, we notice that at $\Lambda_{\rm TMM}  $ where initial values i.e., $\theta_{23} = 45^\circ, ~\rho = \sigma = 90^\circ  $ have been utilized under the symmetry,  
the leading order contributions vanishes due to the presence of $ (c^2_{23} - s^2_{23})$ and $ \sin2\rho, \sin2\sigma $ terms, respectively. Hence, one excepts very mild contributions due to RGE triggered symmetry breaking effect. This has been identified in our numerical study as shown by the first panel of the bottom row of Fig.~\ref{fig:RGEplot}. 

In the fourth panel, we show the impact of RGE induced symmetry breaking effect on the neutrinoless double beta ($ 0\nu\beta\beta $) decay  experiments. The $ 0\nu\beta\beta $ decay $ (A, Z) \longrightarrow (A, Z+2) + 2e^-$ is the unique process which can probes the Majorana nature of massive neutrinos.
%
The experiments that are currently searching for the signature of $ 0\nu\beta\beta $-decay  are GERDA Phase II \cite{Agostini:2018tnm}, CUORE \cite{Alduino:2017ehq}, SuperNEMO \cite{Barabash:2011aa}, KamLAND-Zen \cite{KamLAND-Zen:2016pfg} and EXO \cite{Agostini:2017jim}.
 However, this process violate lepton number by two-units. The half-life of such decay process is given by \cite{Rodejohann:2011mu,*Dev:2013vxa},
\begin{equation}
(T^{0\nu}_{1/2})^{-1} = G_{0\nu}|M_{0\nu}(A,Z)|^{2} |\langle m \rangle_{ee}|^{2} \;,
\end{equation}
where $  G_{0\nu}$ stands for two-body phase-space factor, $ M_{0\nu} $ is the nuclear matrix element (NME). $|\langle m \rangle_{ee}|$ denotes the effective Majorana neutrino mass. Note that now onwards, we use $ |\langle m \rangle_{ee}| = |m_{ee}| $  for simplicity.
The expression of $|\langle m \rangle_{ee}|$ is given by,
\begin{equation}
|m_{ee}| = \left| \sum^3_{i = 1} m_i U^2_{e i} \right| \;,
\end{equation}
where $U$ stands for PMNS mixing matrix as mentioned in Eq.~(\ref{eq:pmns}). In the standard formalism, one can parameterize $ |m_{ee}| $ as
\begin{align}\label{eq:EffMee}
 |m_{ee}| & = | m_1 c^2_{12} c^2_{13} e^{ 2 i \rho} + m_2 s^2_{12} c^2_{13} e^{ 2 i \sigma} + m_3 s^2_{13} e^{- 2 i \delta} | \;, \nonumber \\
\end{align}
where  $c^{}_{ij} (s^{}_{ij})$ are the leptonic mixing angles, $ \delta,$ and $\rho, \sigma$ stand for the Dirac, Majorana CP-phases, respectively.
%
Also, as we have  information about the mass-squared differences $ \Delta m^2_{21} $, $  \Delta m^2_{31}$ from the neutrino oscillation data, one defines masses $ m_2 $, $ m_3 $ in terms of the lightest neutrino mass $m_1$ as $ m_2 =\sqrt{m^2_1 + \Delta m^2_{21}}$ and $ m_3 =\sqrt{m^2_1 + \Delta m^2_{21} + \Delta m^2_{31}} $ for the normal mass ordering.
%

The last panel of Fig.~\ref{fig:RGEplot} shows the behavior of $  |m_{ee}| $. Using the latest $ 3\sigma $ data of oscillation parameters, we present the allowed area by the light-gray color within the dotted-black lines.
The RGE-induced breaking pattern within the \tmm~  has been shown by the brown color \footnote{Recently, the pattern of $  |m_{ee}| $ due to RGE corrections within the framework of $ \mu-\tau $ reflection symmetry have been discussed in ~\cite{Nath:2018hjx, *Nath:2018zoi}.}.
 The most stringent upper limit on the effective Majorana neutrino mass $  |m_{ee}| $ arises from KamLAND-Zen experiment \cite{KamLAND-Zen:2016pfg}. Their collaboration  have recently reported the bound on $  |m_{ee}|  < (0.061 - 0.165)$ eV  at 90\% C.L. by taking into account the uncertainty in the estimation of the  nuclear matrix elements, which we show by the horizontal blue band in the fourth panel. 
 On the other hand, the upper bound for the lightest neutrino mass is shown by the vertical blue band. This can be read from  the recent  \textit{Planck} report~\cite{Aghanim:2018eyx} which gives $ \sum m_{\nu} < 0.12$ eV (95\%, \textit{Planck} TT, TE, EE + lowE + lensing + BAO).
%
One can define $ |m_{ee}|  $ at $ \Lambda_{\rm TMM} $ within the underlying symmetry by inserting  initial values of phases $ \delta, \rho$  and  $\sigma $ as
\begin{align}\label{eq:EffMeeAtHE}
 |m_{ee}| (\Lambda_{\rm TMM}) & = | m_1 c^2_{12} c^2_{13}  + m_2 s^2_{12} c^2_{13}  + m_3 s^2_{13}| \;; \quad {\rm for } ~ -\delta = \rho = \sigma = 90^\circ \;.
\end{align}

We notice from Eq.~\ref{eq:EffMee} that the cancellation among the various terms of  $ |m_{ee}|  $ depends on the CP-phases. Thus, their  breaking patterns at low energies play very important role to understand the numerical results. One sees from  Fig.~\ref{fig:RGEplot} that all the CP-violating phases show less than $ \mathcal{O}(1^\circ) $ deviation from their symmetry limit. This tells that there can not be any significant cancellations among the different terms of $ |m_{ee}|  $ at low energies.
This is apparent from Eq.~\ref{eq:EffMeeAtHE} i.e., some components can not attain negative sign with less than $ \mathcal{O}(1^\circ) $ deviation, whence there will not be proper cancellations. 
From the last panel of Fig.~\ref{fig:RGEplot}, we notice that minimum of $ |m_{ee}| $  never approaches to zero.
We find that $|m_{ee}|$ can reach $\sim 2 $ meV for $m_1 \rightarrow 0.1 $ meV. On the other hand,  the upper limits of  $|m_{ee}|$  can be $ \sim 35 $ meV.
%
%
\section{Explicit Breaking of TMM}\label{sec:ExpBrm}
As observed in previous section that the RGE-induced breaking effect significantly explains the latest best-fit value of the mixing parameter $\theta_{12}^{}$~\cite{Esteban:2018azc}. However, the deviations for $\theta_{13}^{}, \theta_{23}^{}$ and CP-phases are very mild and less than  $ \mathcal{O}(1^\circ) $ have been observed, which are much smaller than their latest best-fit values \cite{Esteban:2018azc}. Although such small deviations are in compatible with current $ 3\sigma  $ experimental data, it may become necessary to consider large deviations when more accurate data will be included. In this section, we discuss the breaking of \tmm~ to explain the low energy data by introducing explicit breaking terms in the neutrino mixing matrix.

We introduce explicit breaking terms in such a way that one can explain large $\theta_{12}^{}$ as well as non-maximal values of $ \theta_{23}^{} $ and $ \delta $, which are in well agreement with the latest global analysis data~\cite{Esteban:2018azc,Capozzi:2016rtj,deSalas:2017kay}. Moreover, although the predicted $ \theta_{13}^{} $ within this mixing pattern is in compatible with the latest $ 3\sigma $ data, we show that such explicit breaking can also explain its best-fit value.
Introducing breaking terms in $ 1$-$2 $, $ 2$-$3 $ rotation matrices, one can re-write Eq.~\ref{eq:VtmmRot} as
\begin{equation}\label{eq:Vprime}
V^{\prime} = P_l \otimes O_{23}\left( \dfrac{\pi}{4}, \dfrac{\pi}{2} + \delta_{\epsilon}\right)  \otimes O_{13}\left( \dfrac{\pi}{4},0\right)  \otimes
O_{12}\left( \dfrac{\pi}{4} + \epsilon_{12},0\right)  \otimes O_{13}\left( \dfrac{\pi}{4},\pi\right)  \; ,
\end{equation}
where $ \delta_{\epsilon}, \epsilon_{12} $ are the explicit breaking parameters.
Expanding Eq.~\ref{eq:Vprime} upto leading order terms, we get 
\begin{eqnarray}\label{eq:BreakTmm}
V^{\prime} = V_0 & + & \frac{1}{4} \epsilon_{12}
\left(\begin{matrix}-\sqrt{2} & 2 & \sqrt{2} \cr -\sqrt{2} - i & -2
+ i\sqrt{2} & \sqrt{2} + i \cr -\sqrt{2} + i & -2
- i\sqrt{2} & \sqrt{2} - i \end{matrix}\right)+
 \frac{1}{4} \delta_{\epsilon}
 \left(\begin{matrix} 0 & 0 & 0 \cr
    1 - \sqrt{2} & \sqrt{2} &  - 1 - \sqrt{2} \cr
     i \sqrt{2} &   i 2 &   i \sqrt{2}  
     \end{matrix}\right) \nonumber \\
& + & 
{\cal O}(\epsilon^2_{13},\delta^2_{\epsilon}, \epsilon_{13}\delta_{\epsilon} )\; .
\end{eqnarray}
It is  clear from Eq.\ref{eq:BreakTmm} that if one considers perturbative terms $ \epsilon_{12} \neq 0, \delta_{\epsilon} = 0$ in such scenario Eq.\ref{eq:BreakTmm} does not serve our intention to explain large $\theta_{12}^{}$ together with non-maximal $ \theta_{23}^{}$, and $\delta $. As in absence of $ \delta_{\epsilon}$, one can see that  $ |V^{\prime}_{\mu i}| = |V^{\prime}_{\tau i}| $ is still maintained and hence maximality remain preserved for $ \theta_{23}^{}$, and $\delta $. However, non-zero $ \delta_{\epsilon}$ breaks this equality and able to give non-maximal values of $ \theta_{23}^{}, \delta $.
In order to have better understanding, we calculate mixing angles and Jarlskog invariant $  J_{CP}$ 
as

\begin{eqnarray}
\sin^{2}\theta_{13} & = & \dfrac{1}{8} (3-2\sqrt{2})+\dfrac{1}{4} (\sqrt{2}-1)\epsilon_{12} + 
 {\cal O}(\epsilon^2_{12})\nonumber \;, \\
\sin^{2}\theta_{12} & = & \dfrac{2}{5+2\sqrt{2}}+\dfrac{4}{289} (12+19\sqrt{2})\epsilon_{12} + 
 {\cal O}(\epsilon^2_{12})\nonumber \;, \\
\sin^{2}\theta_{23} & = & \dfrac{1}{2} - \dfrac{1}{17} (6 + \sqrt{2})\sin \delta_{\epsilon} -
 \dfrac{2}{289} (45 - \sqrt{2})\epsilon_{12} \sin \delta_{\epsilon} + {\cal O}(\epsilon^2_{12},\delta^2_{\epsilon})\nonumber \;, \\
 J_{CP} & = & - \dfrac{1}{32} (1+ 3\epsilon_{12} + \dfrac{3}{2}\epsilon^2_{12} -  \dfrac{1}{2}\delta^2_{\epsilon} )   + {\cal O}(\epsilon^2_{12} \delta^2_{\epsilon})\;.
\end{eqnarray}
Note that in calculating last two expressions, we kept next-to-leading order terms. 

\begin{figure}[h!]
\centering
\includegraphics[height=8cm,width=12cm]{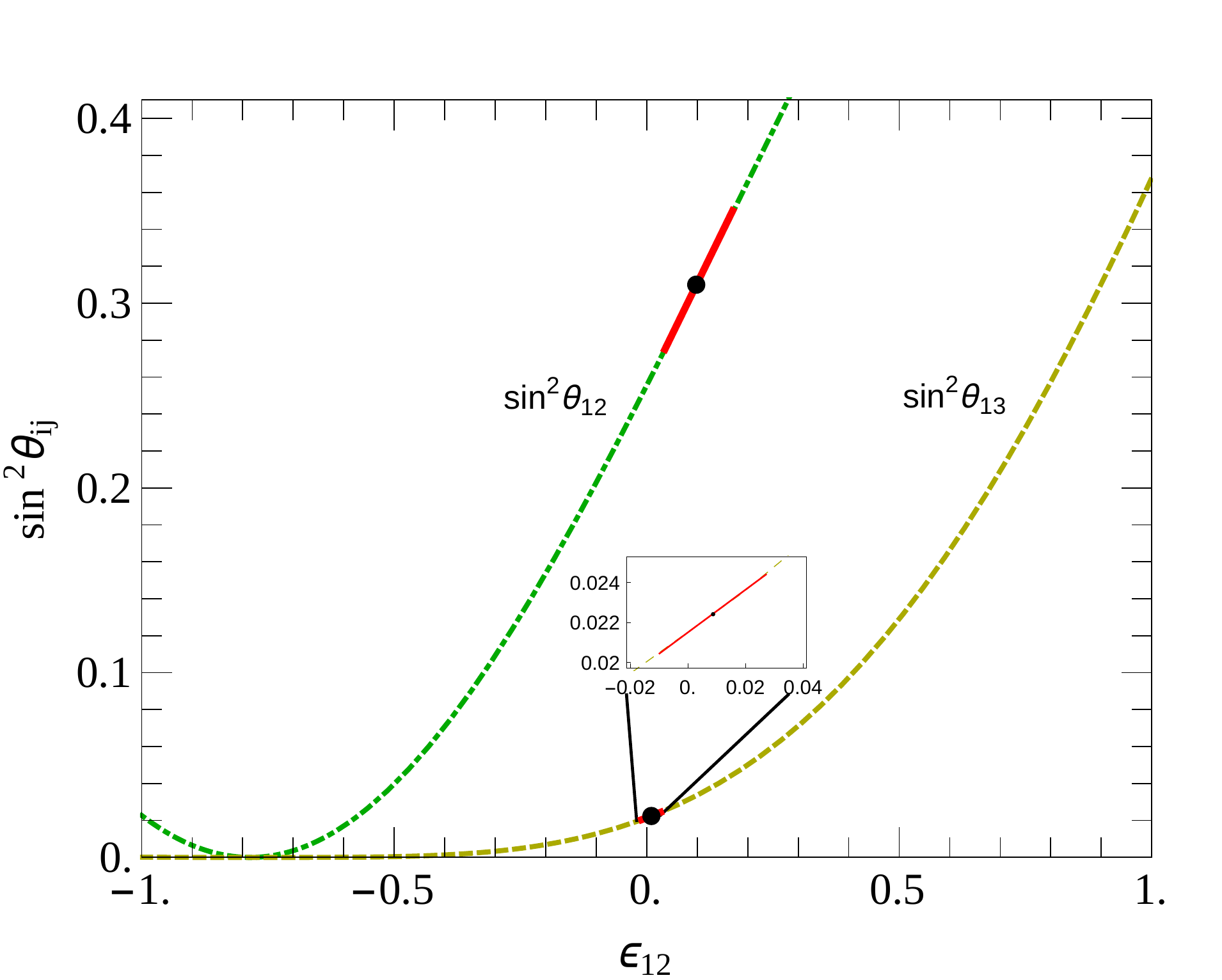}
\caption{\footnotesize 
Correlations between $\sin^2 \theta_{12}, \sin^2 \theta_{13}$ vs $\epsilon_{12}  $ are shown by the dark green (dash-dotted ) and light green (dashed) lines, respectively. The red color patch and black-dot represent 3$ \sigma $ allowed region and best-fit value of the latest data, respectively.}
\label{fig:sij-eps12}
\end{figure}

From Fig.~\ref{fig:sij-eps12}, we notice that an infinitesimal value of $\epsilon_{12}  $ is able to explain the whole 3$\sigma$ range of the mixing angle $\theta_{13}$ (see the dashed light green curve). The latest 3$\sigma$ range of $\theta_{13}$ is shown by the red color patch, where as the best-fit value is marked by the black dot.
The dashed-dotted dark green curve shows our numerical results for  $\theta_{12}$. One can see that $\epsilon_{12} \sim 0.1$ is able to generate the latest best-fit value of  $\theta_{12}$, which is marked by the black dot, whereas $\epsilon_{12} \sim 0.18$ is able to explain the current 3$\sigma$ range of $\theta_{12}$ as shown by the red patch.

\begin{figure}[h!]
\centering
\includegraphics[height=8cm,width=18cm]{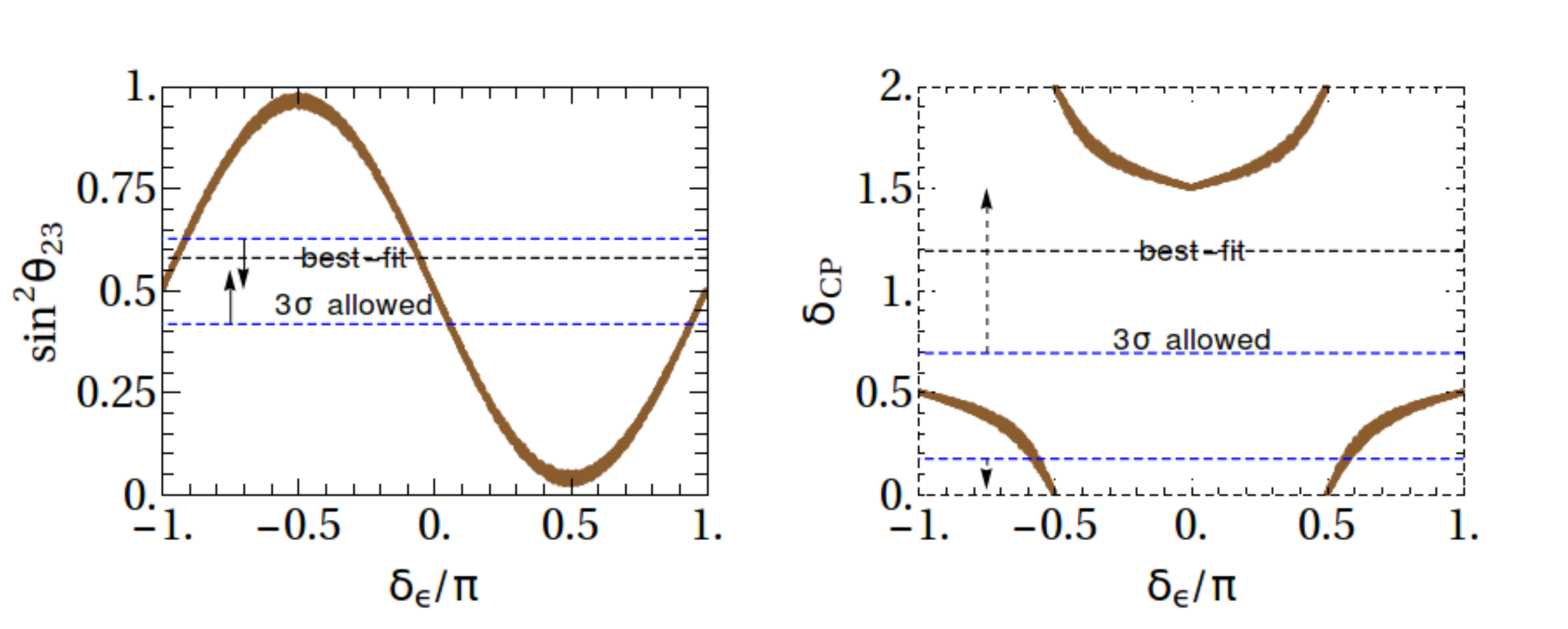}
\caption{\footnotesize 
Left: Correlation between $\sin^2 \theta_{23}$ vs $\delta_{\epsilon} $ as shown by the brown curve. Right: Correlation between $\delta_{CP}$ vs $\delta_{\epsilon} $ as shown by the brown curve. The black dash and the blue dash lines show the latest best-fit value and the  3$\sigma$ range, respectively.}
\label{fig:eps23-eps12}
\end{figure}

We shown the impact of explicit breaking terms on the oscillation parameters $ \theta_{23}, $ and $ \delta  $ in Fig.~\ref{fig:eps23-eps12}. 
As we know that any non-zero values of $\delta_{\epsilon} $ will lead to non-maximal values of $ \theta_{23}, $ and $ \delta  $,   which is apparent from the figure. The best-fit and  the 3$\sigma$ range of the latest global analysis of neutrino oscillation data are shown by the black dash and the blue dash lines, respectively. 
We notice from the left panel figure that an infinitesimal value of $\delta_{\epsilon} $ is able to generate the latest best-fit value of $\theta_{23}$ (see the intersection point between the brown curve and the blue dash curve). On the other hand, it can be seen from the right panel figure that any value of  $\delta_{\epsilon} $ are unable to explains the latest best-fit value of $\delta_{CP}$. However, this explicit breaking term can  explain the values of $\delta_{CP}$ which lies within the current 
3$\sigma$ data.

\section{Summary}\label{sec:summary}
We study here tetra-maximal neutrino mixing pattern, which was originally proposed in~\cite{Xing:2008ie}, to address the latest  global analysis of neutrino oscillation data. This symmetry predicts with $\theta_{12} \approx 30.4^\circ$,
$\theta_{13} \approx 8.4^\circ$, $\theta_{23} = 45^\circ$ and $ - \delta = \rho = \sigma = 90^\circ$. 
According to the latest global analysis of neutrino oscillation data~\cite{Esteban:2018azc,Capozzi:2016rtj,deSalas:2017kay}, one can notice relatively higher best-fit value of $ \theta_{12} $ as well as non-maximal values of both $ \theta_{23}, $ and $ \delta $.
Besides this, though the latest T2K results are in good agreement with the symmetry, the current NO$\nu$A results seem to favor non-maximal $\theta_{23}^{}, $ and $ \delta$.
In order to explain the realistic data, we study the breaking of TMM pattern.
Here we examine the status of tetra-maximal neutrino mixing pattern  for the leptonic mixing matrix by confronting them with current neutrino oscillation data. To do that we first examine the breaking of the symmetry due to renormalization group running effects and then study the impact of explicit breaking term. 

From the RGE-induced symmetry breaking we find less than $\mathcal{O} (1^\circ)$ deviations for the mixing parameters $\theta_{13}, \theta_{23}$ from their predicted values. A significantly large deviation of $\theta_{12}$ has been identified. These numerical results are in good agreement with the analytical expressions as described in Fig.~\ref{fig:sij-eps12}. 
Besides this, for all the three CP-phases less than $\mathcal{O} (1^\circ)$ deviations have been observed. We also study the impact of RGE-induced symmetry breaking on the effective Majorana neutrino mass matrix, $|\langle m \rangle_{ee}|$. It has been found that minimum of $|\langle m \rangle_{ee}|$ can never reach to zero in this framework.  
We also notice that $|m_{ee}|$ can reach $\sim 2 $ meV for $m_1 \rightarrow 0.1 $ meV and the upper limits of  $|m_{ee}|$  can be as large as $ \sim 35 $ meV.
Finally, we discuss the breaking of the symmetry in presence of explicit breaking terms. In this approach, we are able to generate large deviations for the leptonic flavor mixing  parameters $\theta_{13}, \theta_{23}$ and $\delta$ for a relatively small values of the breaking terms. We summarize these results in Fig.~\ref{fig:eps23-eps12}.

\section{Acknowledgements}
%
%
Author is supported by the postdoctoral fellowship program DGAPA-UNAM.
This work is also supported by the  grants  CONACYT CB-2017-2018/A1-S-13051 (M\'exico) and  the  German-Mexican  research  collaboration grant SP 778/4-1 (DFG) and 278017 (CONACYT).

\bibliography{mu-tau}
\end{document}